\documentclass[oldversion]{aa}
\usepackage{txfonts}
\usepackage{graphicx}
\usepackage{natbib}
\bibpunct{(}{)}{;}{a}{}{,}
\begin{document}

\title{Discovery of a rapidly pulsating subdwarf B star candidate in \\ $\omega$ Cen \thanks{Based on observations collected at the European Organisation for Astronomical Research in the Southern Hemisphere, Chile (proposal ID 081.D-0746).}
}
\author{
S.K. Randall \inst{1}
\and A. Calamida \inst{1}
\and G. Bono \inst{2,3}
}

\institute{
ESO, Karl-Schwarzschild-Str. 2, 85748 Garching bei M\"unchen, Germany; \email{srandall@eso.org,acalamid@eso.org}
\and Istituto Nazionale de Astrofisica, Osservatorio Astronomico di Roma, Via Frascati 33, 00040 Monte Porzio Catone, Italy; \email{bono@mporzio.astro.it}
\and Universit\`a di Roma ``Tor Vergata", Department of Physics, Via della Ricerca Scientifica 1, 00133, Rome, Italy
}
\date{Received date / Accepted date}

\abstract
{We report the discovery of the first variable extreme horizontal branch star in a globular cluster ($\omega$ Cen). The oscillation uncovered has a period of 114 s and an amplitude of 32 mmags. A comparison between horizontal branch models and observed optical colours indicates an effective temperature of 31,500$\pm$6,300 K for this star, placing it within the instability strip for rapidly oscillating B subdwarfs.  The time scale and amplitude of the pulsation detected are also in line with what is expected for this type of variable, thus strengthening the case for the discovery of a new subdwarf B pulsator. 
}

\titlerunning{A rapid sdB pulsator in $\omega$ Cen}
\authorrunning{Randall, Calamida \& Bono}
\maketitle

\section{INTRODUCTION}    

Subdwarf B (sdB) stars are hot (20,000 K $\lesssim T_{\rm eff} \lesssim$ 40,000 K), compact (5.2 $\lesssim \log{g} \lesssim$ 6.2), evolved objects found on the extreme horizontal branch (EHB), a high temperature extension of the horizontal branch \citep{heber2008}. Over the last decade, a small subset ($\sim$ 5\%) of them have been found to exhibit rapid luminosity variations with typical periods of 100$-$200 s and amplitudes on the mmag scale (see e.g. \citealt{fontaine2006} for a recent review). Also known as EC 14026 stars after the prototype \citep{kilkenny1997}, the rapid pulsators are found among the hotter sdB stars in a well-defined instability strip from 29,000$-$36,000 K. Pulsation driving is believed to occur through the action of a classical $\kappa$-mechanism associated with a local overabundance of iron peak elements in the driving region, which in turn relies on radiative levitation processes \citep{charp1996,charp1997}. The modelling of these stars has been very successful, to the point where their non-adiabatic pulsation properties are understood not only qualitatively, but also quantitatively. Indeed, full asteroseismological analyses leading to a precise determination of the targets' fundamental parameters have now been carried out for 12 out of at least 35 known EC 14026 stars (see e.g. \citealt{val2008, charp2008} for recent results). 

It is generally accepted that sdB stars are the progeny of red giant branch stars that underwent significant mass loss around the time of the core helium flash, leaving them with very thin hydrogen-rich envelopes. These stellar structures do not experience the asymptotic giant branch phase, instead evolving as so-called {\em AGB-Manqu\'e} stars \citep{greggio1990}, however we still lack firm constraints on the evolutionary channel(s) followed. A number of different scenarios involving e.g. a common envelope phase, stable and unstable Roche lobe overflow, or the merger of two He white dwarfs have been proposed \citep{han2002,han2003}, but remain to be tested observationally. More recently, it has been suggested \citep{han2008} that the low fraction of close binaries identified among cluster EHB stars ($\approx$ 4\% from \citealt{monibidin2008}) compared to the field population ($\approx$ 60\% according to \citealt{maxted2001}; $\approx$ 40\% from \citealt{napi2006}) can be explained naturally in terms of an age effect. In the case of an older population, the white dwarf merger scenario forming a single sdB star becomes more important than common envelope evolution producing an sdB star in a close binary. It has also been suggested  that cluster EHB stars might be the progeny of the ``hot helium-flasher" scenario (\citealt{castellani1993, dcruz1996,castellani2006, miller2008}). In this scenario, red giant stars approaching the tip of the RGB lose a significant fraction of their envelope due to a violent mass-loss event. The ensuing red giants have a higher mass than the limit for core helium ignition, and undergo a He-core flash either when approaching, or along the white dwarf cooling sequence \citep{brown2001}. Another working hypothesis is that cluster EHB stars are the progeny of the helium-enhanced sub-populations \citep{dantona2002,lee2005,dantona2007} recently detected in a few globular clusters \citep{bedin2004,piotto2008}. 

This is where asteroseismology becomes important: since the different evolutionary scenarios leave an imprint on the fundamental parameters of the resulting sdB population, contrasting the predictions from population synthesis models with the corresponding asteroseismic parameter distributions can be used to distinguish between the proposed formation channels. First attempts in this direction have been made \citep{fontaine2008}, however the number of EC 14026 stars analysed up to now is too small for quantitative conclusions.  Moreover, the sample of known sdB pulsators has so far been confined to the Galactic field population. While several attempts have been made to detect variable EHB stars in globular clusters (e.g. \citealt{catelan2008,kaluzny2008,reed2006}), no convincing EC 14026 star candidates were uncovered. 

Rapidly pulsating subdwarf B stars in globular clusters are of high interest in the context of an asteroseismic characterisation of the EHB population in these systems. The first necessary step in this direction is obviously the detection of pulsating subdwarf B stars in globular clusters. This is why, when given a couple of free hours at the end of a night of observations dedicated to a different programme, we decided to monitor a field in $\omega$ Cen for rapid variability among the EHB population. We present the outcome of this mini-survey in what follows.

\section{OBSERVATIONS}

\begin{figure}[t]
\centering
\includegraphics[width=8.8cm,bb=80 85 585 585,clip]{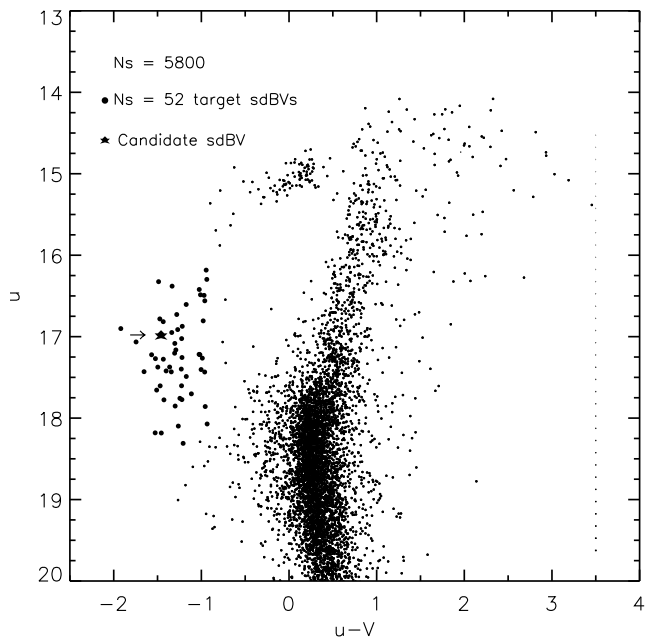}
\caption{CMD of $\omega$ Cen based on our SUSI2 $u$-band data combined with the WFI $V$-band data. The 52 selected subdwarf B star candidates are marked by large black points, and the position of the pulsating sdB star candidate is indicated. Error bars on the right display the mean colour and magnitude errors as a function of brightness.}
\label{cmdsusi}
\end{figure}

We obtained 2 hours of time-series photometry of a 5.5$\arcmin$ $\times$ 5.5$\arcmin$ field in the southeast quadrant of $\omega$ Cen with the SUperb Seeing Imager (SUSI2) mounted on the NTT at La Silla, Chile, on 14 March 2008. A $U$-band filter was used to minimize field crowding, and 3$\times$3 binning was chosen in order to reduce the overhead time to 16 s. Combined with an exposure time of 20 s, this resulted in a cycle time of 36 s, short enough to detect the rapid oscillations expected in EC 14026 type stars.

After bias- and flatfield-correcting the images, photometry was performed using DAOPHOT/ALLSTAR and ALLFRAME \citep{stetson1987,stetson1991,stetson1994}. Assuming a Moffat analytical function and using the task ALLSTAR, we first estimated an analytical point-spread function (PSF) for each frame by selecting about 50 bright, isolated stars uniformly distributed over the chip. In order to obtain a global star catalogue where all images reside in the same coordinate system, we used the task DAOMATCH/DAOMASTER \citep{stetson1994} to put together the 192 images obtained for the field. As a reference catalogue, we employed UBVI-band photometry of $\omega$ Cen gathered with the Wide Field Imager (WFI) on the 2.2-m ESO/MPI telescope at La Silla \citep{castellani2007}, which completely overlaps the SUSI2 field. Finally, we performed simultaneous PSF-fitting photometry on the SUSI2 data frames using ALLFRAME. The seeing on the individual images ranged from 0.76$\arcsec$ to 1.3$\arcsec$, corresponding to FWHM values between 2.9 and 5.4 pixels, while the photometric accuracy at $u\sim$ 18.5 mag ranges from $\sim$ 0.1 to $\sim$ 0.15 mag. The final star catalogue of the field monitored features $\sim$20,000 stars, of which 52 were selected as subdwarf B star candidates on the basis of brightness (16 $\leq u \leq$ 18.5), colour ($-$2 $\leq (u-V) \leq$ $-$0.8), photometric accuracy $([\delta u,\delta V]_{\rm mean} \leq$ 0.05 mag), sharpness (s $\leq$ 1), and separation index (sep$_{u,V}\geq$ 1\footnote{The separation index quantifies the degree of crowding \citep{stetson2003}.}). In Fig. \ref{cmdsusi} we show the colour magnitude diagram (CMD) obtained by cross-correlating our SUSI2 $u$-band data with the corresponding $V$-band measurements from the WFI catalogue. Note that the data are not calibrated in magnitude; this is however of no relevance for the present study since we are interested only in the {\it relative} brightness variation of the targets.

\section{RESULTS}

\begin{figure*}
\begin{tabular}{cc}
{\includegraphics[width=8.8cm,bb=68 136 512 613,clip]{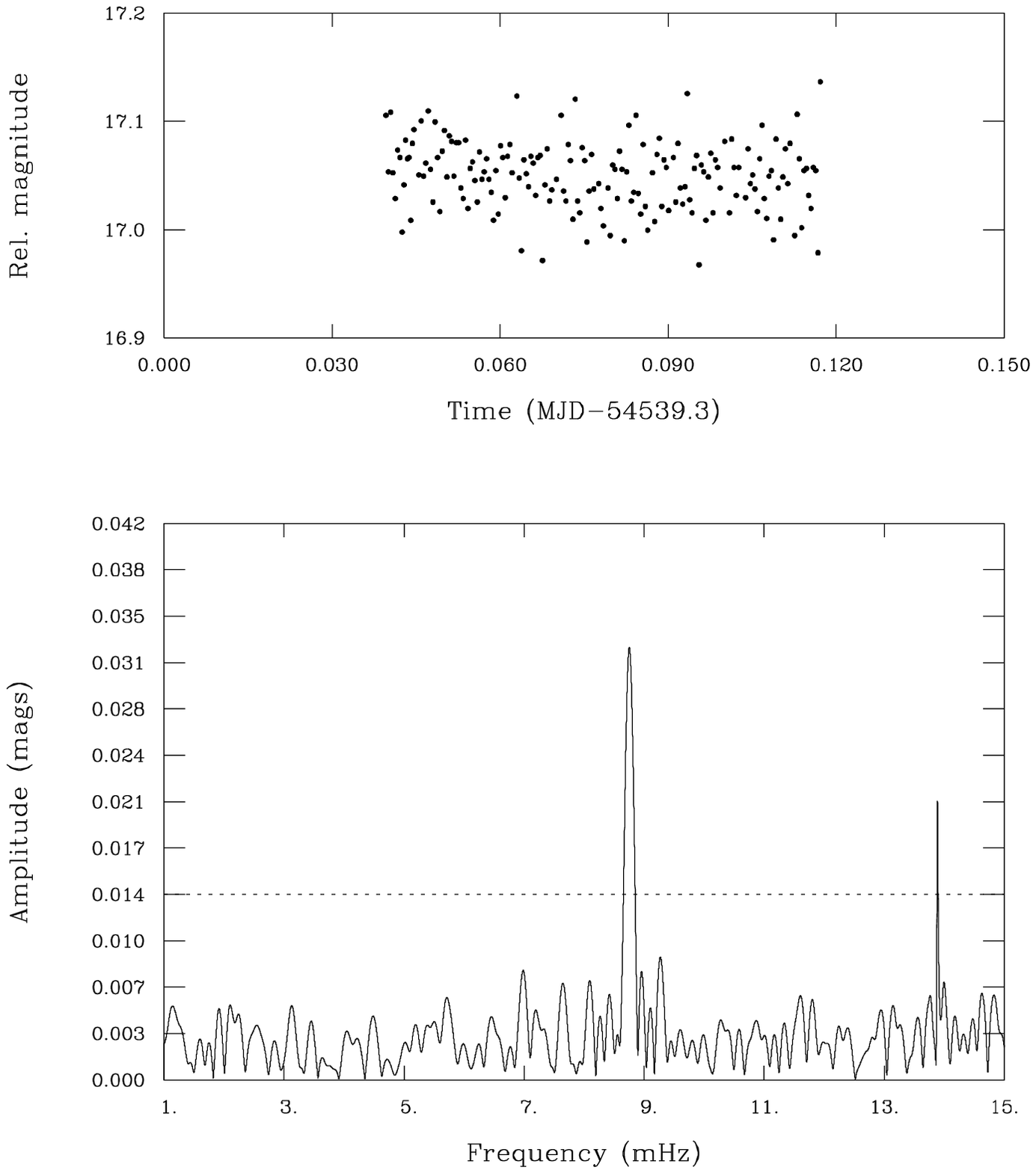}} & {\includegraphics[width=8.8cm,bb=68 136 512 613,clip]{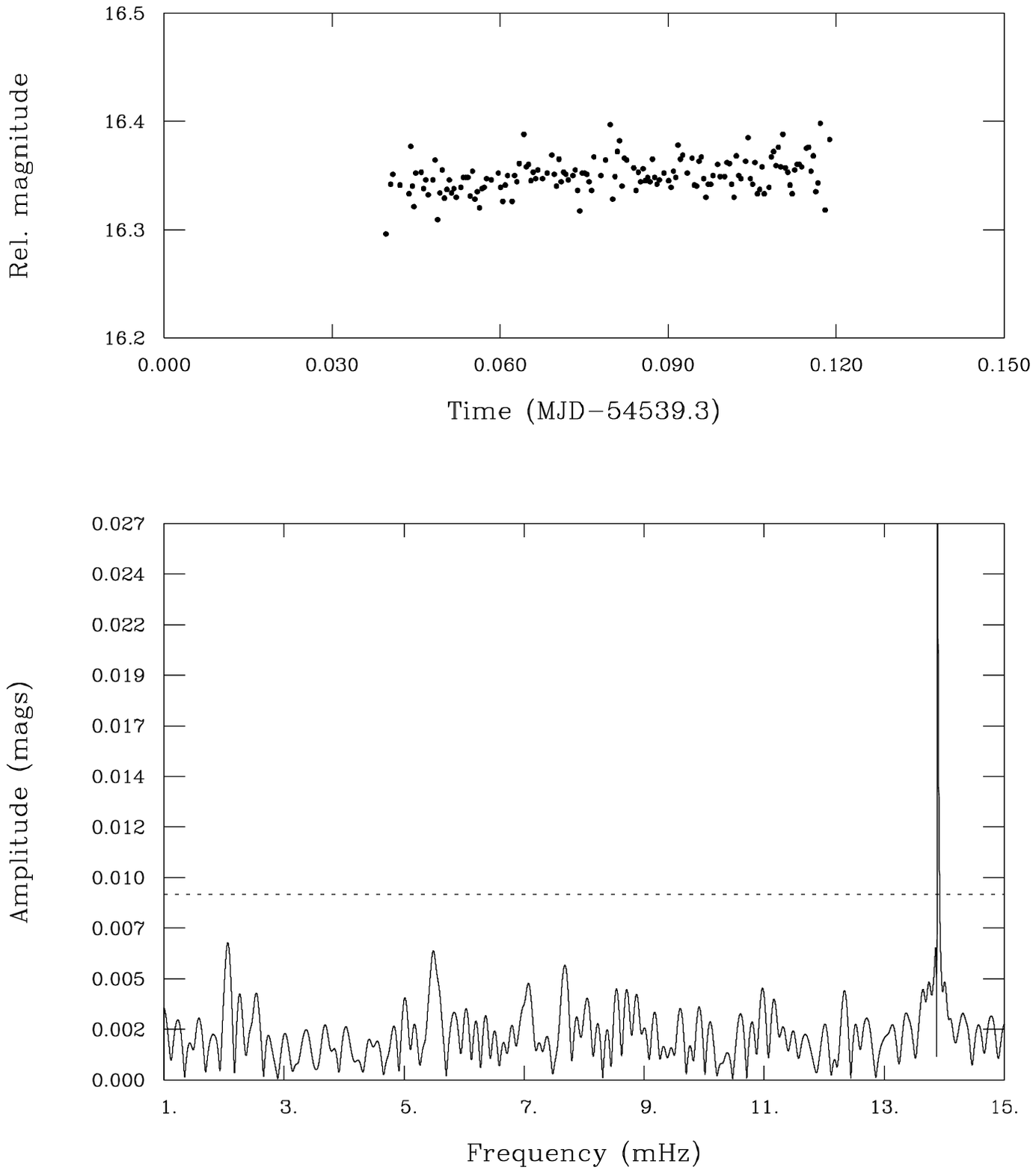}} \\
\end{tabular}
\caption{\it{Left Panel:} \rm Light curve (top) and Fourier transform (bottom) for the pulsating EHB star discovered in $\omega$ Cen. The relative $u$ magnitude indicated for the light curves is offset by around +0.25 mags with respect to the absolute magnitude from the WFI catalogue. The horizontal dashed line indicates the 4$\sigma$ detection threshold. \it{Right Panel}: \rm The same as in the left panel, but for a comparison star deemed to be constant.}
\label{fts}
\end{figure*}

\begin{figure}[b]
\centering
\includegraphics[width=8.0cm]{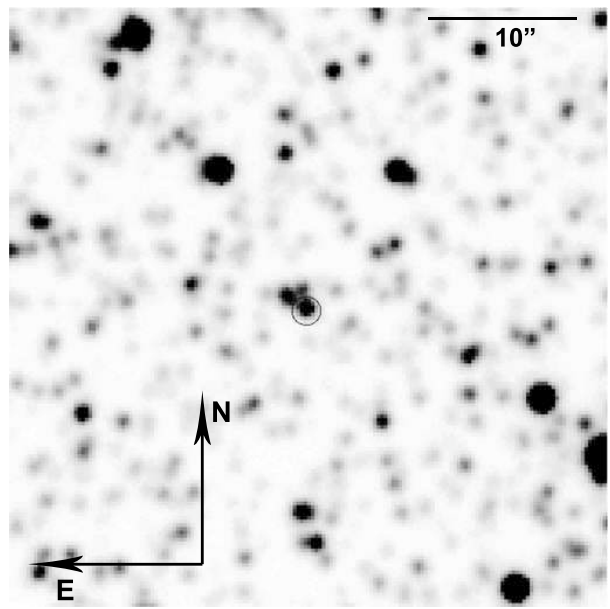}
\caption{Finding chart for the variable EHB star discovered in $\omega$ Cen.}
\label{fc}
\end{figure}

\begin{figure}[h]
\centering
\includegraphics[width=8.0cm,bb=63 233 545 620]{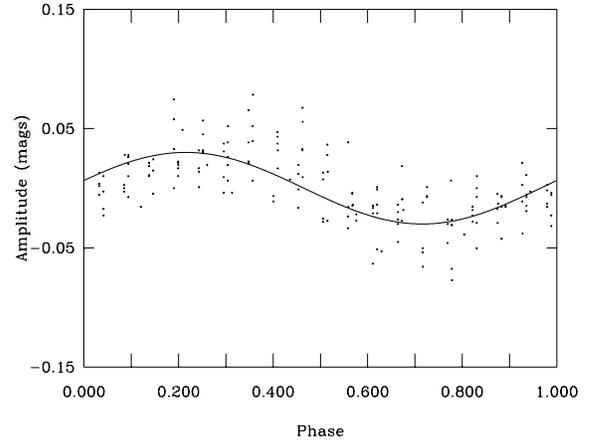}
\caption{Phase-folded light curve for the variable EHB star, based on the 114-s periodicity detected. The sinusoid overplotted constitutes a least squares fit to the light curve and has an amplitude of 0.03 mags and a phase of 0.22.}
\label{phasefolded}
\end{figure}

Selecting only the 52 EHB stars indicated in Fig. \ref{cmdsusi}, we computed the light curves with respect to the mean $u$-band magnitude, incorporating corrections for airmass and seeing variations by scaling the SUSI2 magnitudes to a WFI $U$-band reference magnitude. Fourier transforms were then calculated for each light curve in the 1$-$15 mHz range, appropriate for detecting the pulsations expected for the EC 14026 pulsators. We display two selected pairs of light curves and the corresponding Fourier transforms in Fig. \ref{fts}. The lefthand plot is particularly interesting since it features a strong peak well above the adopted 4$\sigma$ detection threshold at a period of 114 s (8.75 mHz) and an amplitude of $\sim$30 mmag, which is compatible with EC 14026-type pulsations. The secondary peak is clearly an artefact of the data as it corresponds to the Nyquist frequency (or a period of 72 s, twice the sampling time), and is encountered in the Fourier transform of most of the targets monitored. On the other hand, the stronger 114s luminosity variation is unique to the star displayed, and it can be seen clearly in the phase folded light curve displayed in Fig. \ref{phasefolded}. The righthand panel of Fig. \ref{fts} features a typical EHB star from our sample that is judged constant on short time scales, with only the 72-s (13.89 mHz) artefact clearly visible. 

The variable star uncovered has J2000.0 coordinates $\alpha$=13$\fh$27$\fm$11.77$\fs$ and $\delta$=$-$47$\degr$32$\arcmin$29.0$\arcsec$. A finding chart based on the combined SUSI2 image is shown in Fig. \ref{fc}. It can be seen that the candidate pulsator is located close to two other stars of similar brightness, with a separation of $\sim$ 1.5 times the average FWHM. In fact, space-based ACS imaging of the same region \citep{castellani2007} reveals a faint companion to our target that is not resolved by SUSI2 and is therefore included in the PSF computed for the target. However, the flux contribution of this companion is only $\sim$1\%, too small to potentially induce the $\gtrsim$3\% luminosity variation detected. A verification of the photometry shows no periodic change in brightness for the stars closest to our pulsator candidate, and moreover the residuals after PSF subtraction are negligible. We therefore believe that the 114-s luminosity variation observed is inherent to the star itself rather than constituting an observational artefact.

\begin{figure*}[t]
\begin{tabular}{cc}
{\includegraphics[width=8.8cm]{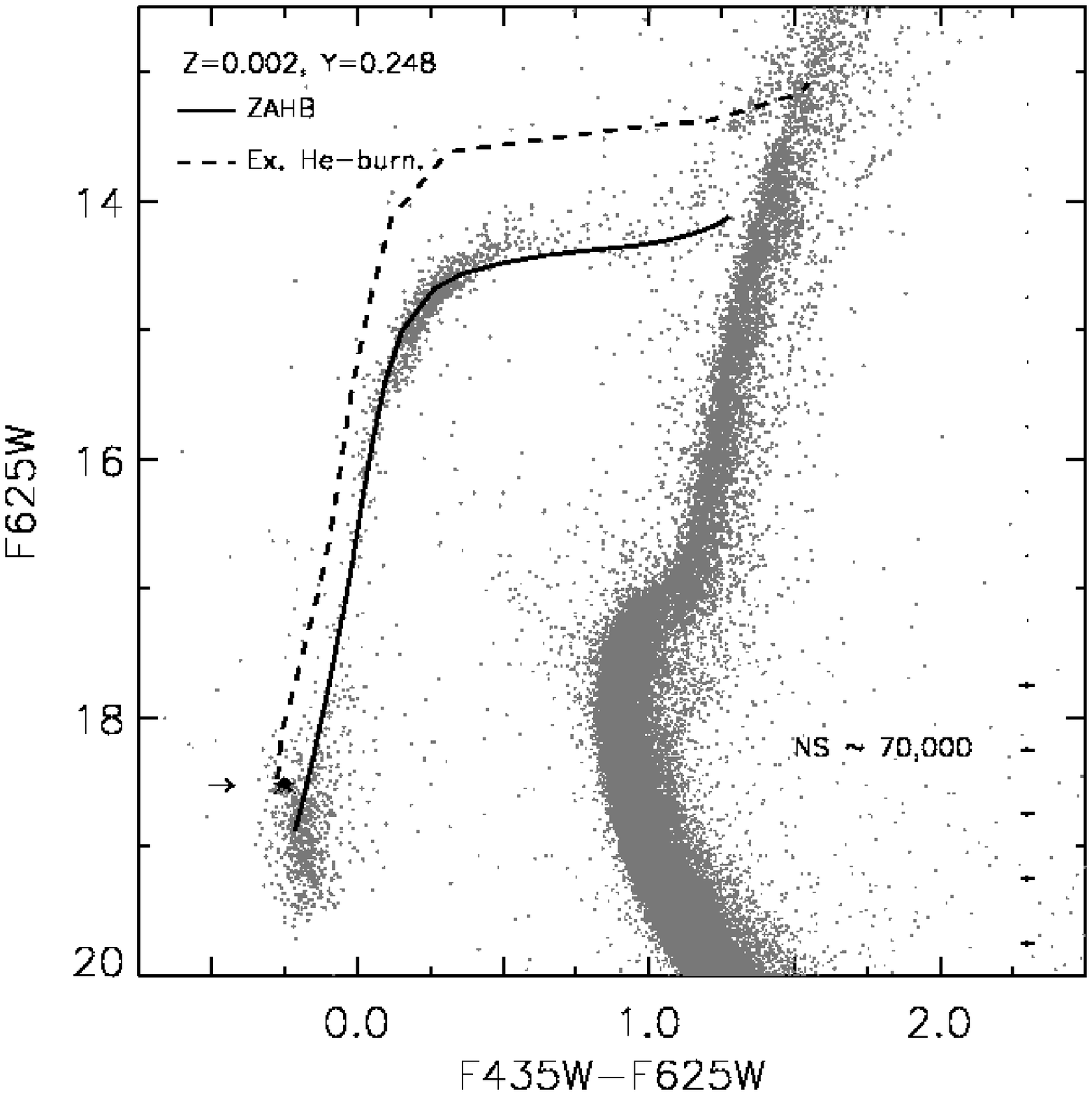}} & {\includegraphics[width=8.8cm]{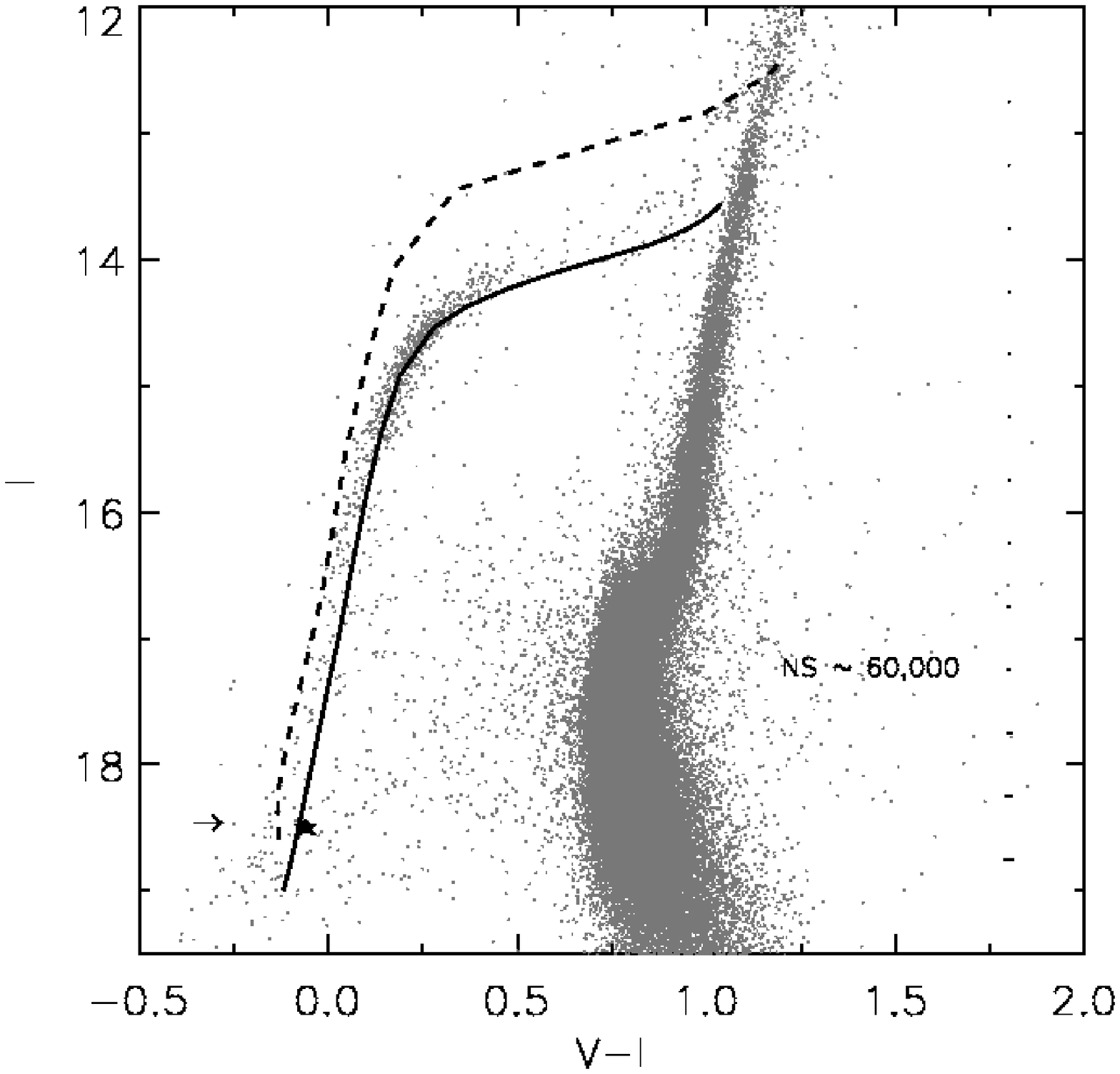}} \\
\end{tabular}
\caption{$F625W, F435W-F625W$ (left) and $I, V-I$ (right) CMDs for selected stars of  the $\omega$ Cen ACS and WFI catalogues.  Details concerning the selection criteria for the star catalogues used in the plot are given in \citet{castellani2007}. The error bars on the right of each plot account for colour and magnitude photometric errors, and the number of stars in each catalogue is indicated. We overplot the ZAHB and the exhaustion of central He burning for the labelled composition. The candidate sdB pulsator is marked by an arrow and asterisk in both plots.}
\label{CMD}
\end{figure*}

Since the SUSI2 field overlaps with both the WFI and the ACS catalogues of $\omega$ Cen, we can determine the magnitudes and colours of the variable star in both filter systems. These are $B_{ACS}$=18.30, $R_{ACS}$=18.55, $H\alpha_{ACS}$=18.48, $U_{WFI}$=16.81, $B_{WFI}$=18.27, $V_{WFI}$=18.46, and $I_{WFI}$=18.52. The location of the pulsator in the CMD obtained on the basis of both the WFI and ACS data is shown in Fig. \ref{CMD}, and reveals this star to be a typical member of the $\omega$ Cen EHB population. We can therefore employ appropriate horizontal branch models in order to derive an estimate of the effective temperature of our target. High S/N spectroscopy would yield more precise results, however we do not have such measurements at our disposal for the time being. In Fig.  \ref{CMD} we have overplotted the zero-age horizontal branch (ZAHB) and the exhaustion of central He-burning publicly available from the BASTI database (see, e.g. \citealt{pietrinferni2004}) with an alpha enhancement [$\alpha$/Fe]=0.4, helium abundance $Y$=0.248, and metallicity $Z$=0.002. We assumed a distance modulus $\mu_0$=13.70$\pm$0.06 \citep{delprincipe2006} and reddening $E(B-V)$=0.11$\pm$0.02 \citep{calamida2005}. It can be seen that in both plots the pulsator is located close to current evolutionary predictions within the theoretical and empirical errors. Given an estimated accuracy of $\sim$ 10 \% for the properties of the horizontal branch (HB) models and  the slight dependency on the metallicity assumed (we averaged values for metallicities of Z=0.002 and Z=0.0006, appropriate for the metal-rich and metal-poor HB populations found in $\omega$ Cen) we conservatively estimate $T_{\rm eff}$ = 31,500 $\pm$ 6300 K for the variable star uncovered. This falls right into the  29,000$-$36,000 K instability strip for rapid sdB pulsators. The mass is estimated to be $\sim$0.5 $M_{\odot}$, which is typical for a subdwarf B star. Given that the period and amplitude of the oscillation are also in line with what is expected for this type of pulsator, we believe that we have uncovered the first EC 14026 star in a globular cluster.

\section{CONCLUSION}

On the basis of 2 hours of SUSI2 time-series data, as well as the availability of WFI and ACS multi-colour photometry, we discovered what appears to be the first rapidly pulsating subdwarf B star in a globular cluster. Both the pulsation properties and the effective temperature estimate are in line with what is observed for this type of variable in the field population; however, a longer time series of photometry and high S/N spectroscopy are needed to verify the results presented here. It will be particularly interesting to see whether more frequencies can be uncovered, since all known sdB pulsators show multi-periodic luminosity variations. However, given the magnitude of the pulsator discovered and the shortness of the time series obtained, it is not surprising that only one periodicity was detected. Indeed, the period spectra of fast sdB pulsators are quite often dominated by one high-amplitude mode, and individual frequencies can be too closely spaced to be distinguishable in just one two-hour run. The amplitude of the oscillation measured is among the highest found for any sdB pulsator (the highest amplitude pulsation is around 50  mmags for Balloon 09010001, see \citealt{oreiro2004}; typical amplitudes are on the 1-5 mmag scale), and lower amplitude luminosity variations will be challenging to uncover given the faintness of the star and the crowded field. Nevertheless, we hope that the results presented here will constitute a first step towards the asteroseismology of globular cluster subdwarf B stars, and create the unique opportunity to provide quantitative constraints on their evolutionary history.

\begin{acknowledgements}
S.K.R. would like to thank the ESO La Silla staff, in particular SUSI2 instrument scientist Alessandro Ederoclite for their support. Sadly, the observations reported here were among the last ever obtained with SUSI2 since the instrument has since been decomissioned.
\end{acknowledgements}

\bibliographystyle{aa}
\bibliography{1266}

\end{document}